\documentclass[a4paper,fleqn,usenatbib]{mnras}

\usepackage{color}

\usepackage{graphicx}	
\usepackage{amsmath}	
\usepackage{amssymb}	
\usepackage[T1]{fontenc}
\usepackage{mathptmx}
\usepackage{txfonts}
\usepackage{ae,aecompl} 

\title[Population statistics of beamed sources-II]{Population statistics of beamed sources. II: Evaluation of Doppler factor estimates.}
\author[Liodakis and Pavlidou]
{I.Liodakis$^{1}$\thanks{liodakis@physics.uoc.gr} and V. Pavlidou$^{1,2}$\\
$^{1}$Department of Physics and ITCP\thanks{Institute for Theoretical
  and Computational Physics, formerly Institute for Plasma Physics}, University of Crete, 71003, Heraklion, Greece\\
$^{2}$Foundation for Research and Technology - Hellas, IESL, Voutes, 7110 Heraklion, Greece\\
}
\begin{document}



\maketitle

\label{firstpage}

\begin{abstract}
In a companion paper we presented a statistical model for the blazar population, consisting of distributions for the unbeamed radio luminosity function and the Lorentz factor distribution of each of the BL Lac and Flat Spectrum Radio Quasar (FSRQ) classes. Our model has been optimized so that it reproduces the MOJAVE distributions of apparent speeds and redshifts when the appropriate flux limit is applied and a uniform distribution of jet viewing angles is assumed for the population. Here we use this model to predict the Doppler factor distribution for various flux-limited samples for which Doppler factors have been estimated in a variety of ways (equipartition, variability + equipartition, inverse Compton dominance) on a blazar-by-blazar basis. By comparing the simulated and data-estimated Doppler factor distributions in each case, we evaluate the different methods of estimating blazar Doppler factors. We find that the variability Doppler factors assuming equipartition are the ones in the best agreement with our statistical model, whereas the inverse Compton Doppler factor method is only suitable for FSRQs. In the case of variability Doppler factors, we find that while random errors are relatively low ($\sim 30\%$), uncertainties are dominated by systematic effects. In the case of inverse Compton Doppler factors, random errors appear to dominate, but are significantly larger ($\sim 60\%$).
\end{abstract}


\section{Introduction}\label{intro}

Blazar observations are shrouded in relativistic effects, due to their preferential alignment of their jets close to our line of sight \citep{Blandford1979}. Decomposing relativistic effects from intrinsic properties would allow us to probe the processes important to jet astrophysics including the jet--black-hole connection, the structure and evolution of jet magnetic fields, the evolution of flaring events in the jet rest frame, and particle acceleration in jets.
  
The jet Doppler factor is a key quantity in any such effort. It is the Doppler factor that determines how much flux densities are boosted and timescales compressed in the observer frame. Additionally,  the Doppler factor is a {\em different} function of the bulk Lorentz factor $\Gamma$ and the viewing angle $\theta$ than the one determining the apparent speeds of jet components. Measurement of both these quantities allows one to solve for both $\Gamma$ and $\theta$. For this reason, measurements of Doppler factors on a blazar-by-blazar basis have been actively pursued. 
Several methods have been proposed, including causality arguments, \citep{Aharonian2007,Jorstad2005,Clausen2013}, emission region geometry \citep{Fan2013,Fan2014}, assumed high-energy emission processes \citep{Ghisellini1993}, and the assumption of equipartition between jet magnetic fields and relativistic electron energy densities \citep{Guijosa1996, Readhead1994, Lahteenmaki1999-III, Hovatta2009}.
However, it is not straight-forward to evaluate the accuracy of such estimates. Often these estimates represent only lower limits to the true jet Doppler factors; in other cases, different methods produce different results for the same sources. Since each method uses several different assumptions that might not hold, it is impossible to determine which method provides the most accurate estimate of the Doppler factor of a source on a blazar by blazar basis.

Here, we take a statistical approach to evaluate the accuracy of various techniques for estimating the Doppler factor of a blazar jet. In a companion paper (\citealp{Liodakis2015}, hereafter Paper I), we presented a population model for each of the BL Lac and Flat Spectrum Radio Quasar (FSRQ) classes of blazars. The model consists of distributions for the intrinsic unbeamed 15 GHz radio luminosity and the Lorentz factor of the jets. Its parameters were optimized using well-measured observable quantities: redshift, and apparent velocity ($\beta_{app}$) as measured by the MOJAVE program (Monitoring Of Jets in Active galactic nuclei with VLBA Experiments, \citealp{Lister2005}). We can use this model in order to derive Doppler factor distributions for any flux-limited sample, independently of variability, flux, or equipartition brightness temperature.

Since our approach is independent of all assumptions entering methods of Doppler factor estimation in individual blazars, and is able to adequately describe blazars as a population, we are presented with a unique opportunity: we can use the derived Doppler factor distributions in order to compare them with those obtained using various single-blazar techniques, and evaluate whether each method of evaluating Doppler factors for individual sources yields consistent results, if not on a blazar-by-blazar basis, at least for the population overall. In this way, we can evaluate whether the assumptions used in each of these methods hold.

This paper is organized as follows. In \S \ref{DFdistr} we describe our
model and the resulting Doppler factor distributions, and in \S \ref{DFest} the various Doppler factor estimation techniques we compare against in the following sections. In \S \ref{comparison} we compare our statistical results with known methods of determining Doppler factors. In \S \ref{discussion} , we discuss the results of this comparison in relation to blazar physics, and we attempt to extract the level of error for Doppler factor estimates in individual blazars, for these techniques that are in overall agreement, at the population level, with our model. In addition, we test whether the derived Doppler factor distributions from our model can be themselves applied to individual sources to extract information about their Doppler factors. We summarize our conclusions in \S \ref{conclusions}.

The cosmology we have adopted throughout this work is $H_0=71$ ${\rm km \, s^{-1} \, Mpc^{-1}}$, $\Omega_m=0.27$ and $\Omega_\Lambda=1-\Omega_m$ \citep{Komatsu2009}. This choice was made so that our cosmological parameters agree with the MOJAVE analysis \citep{Lister2009-2}.

\section{Doppler Factor Distributions}\label{DFdistr}

\begin{figure}
\resizebox{\hsize}{!}{\includegraphics[scale=1]{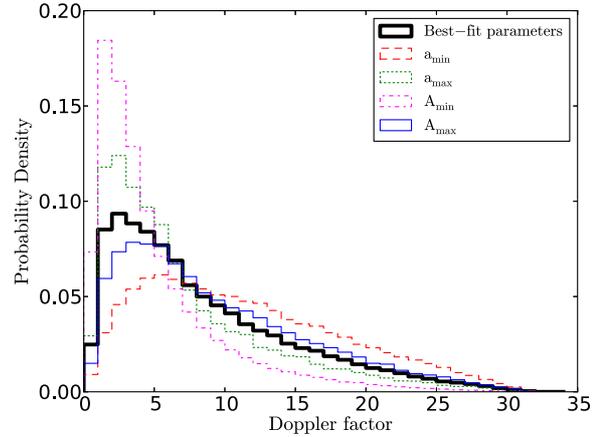}  }
\caption{The Doppler factor distribution for the BL Lac objects with the limits of the model parameters.}
\label{bllim}
\end{figure}

\begin{figure}
\resizebox{\hsize}{!}{\includegraphics[scale=1]{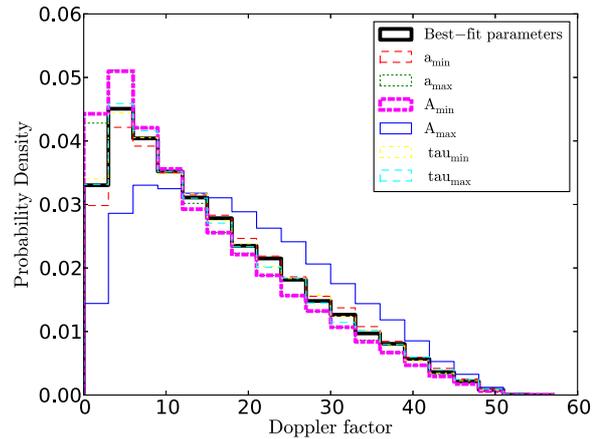}  }
\caption{The Doppler factor distribution for the FSRQs with the limits of the model parameters.}
\label{qsolim}
\end{figure}

In this section we give a short description of our blazar population model. For a  more detailed description see Paper I. 

We optimized our model using data from the MOJAVE survey \citep{Lister2009-2}.  MOJAVE uses a statistically complete flux- limited sample \citep{Arshakian2006}. Samples determined only by strict statistical criteria are crucial for population studies such as ours. We removed outliers and any source that showed unusual behavior (jet bending, inward motions etc.) as indicated by \citet{Lister2009-2, Lister2013}, and separated the sample in two sub-populations: BL Lac objects, and  Flat Spectrum Radio Quasars (FSRQs). A flux limit, at 1.5 Jy, serves as a constraint determining how beamed the set of the observed sources is.

We assumed single power law distributions for the Lorentz factor, 
 \begin{equation}
p(\Gamma)\propto \Gamma^{-\alpha},
\end{equation}
 and the unbeamed luminosity function \citep{Lister1997,Cara2008,Chatterjee2008,Abdo2010},  and have adopted a pure luminosity evolution model described in \citep{Padovanni1992,Padovani1992-2}. 
\begin{equation}\label{lumfunc}
n(L_\nu ,z)\propto \left( \frac{L_\nu }{e^{{T(z)}/\tau}} \right)^{-A}\,, 
\end{equation}
where $T(z)$ is the lookback time. We assumed random viewing angles $\theta$ (i.e. $\cos \theta$ uniformly distributed between 0 and 1). 
Thus, our model parameters are the power law indices for the Lorentz factor distribution ($\alpha$) and the luminosity function ($A$), and the evolution parameter ($\tau$). 

For every set of model parameters, we compared simulated and observed distributions for apparent speeds and redshifts of sources obeying the flux limit. We rejected any model for which the Kolmogorov-Smirnov test (K-S test) delivered a probability $<5\%$ of consistency between observed and simulated distributions. Our optimal model is the one which minimized the product of the K-S probability values for these two distributions. The best fit parameters for the FSRQ population are:$\alpha=0.57\pm0.001$; $A=2.6\pm0.01$; $\tau=0.26\pm0.001$. For the BL Lac population we found:$\alpha=0.738\pm0.002$; $A=2.251\pm0.02$. Note that the BL Lac parameters do not include the evolution parameter: we found that the BL Lac luminosity function is consistent with no evolution. The error represents scanning step and statistical variations in simulated distributions. We have also explored the threshold of the acceptability of our model, by keeping all the parameters but one to the best fit value, and changing the other towards higher or lower values until the K-S test threshold of 5\% is violated. The limits of the parameters are shown in Table \ref{tab:Parameter limits}. 
For each model (determined by a set of $A$, $\alpha$, and $\tau$) and each flux limit value, we can produce distributions of derived quantities, including Doppler factors, viewing angles, and timescale modulation factors. These results  are presented in detail in Paper I along with a detailed discussion on the reasoning behind our assumptions and our optimization algorithm. 

In Figs.~\ref{bllim} and \ref{qsolim}  we review the Doppler factor distributions produced by our model for BL Lacs and FSRQs respectively for a 1.5 Jy - limited sample. 
The Doppler factor is given by
\begin{equation}\label{eqdop}
 \delta=\frac{1}{\Gamma(1-\beta\cos\theta)},
 \end{equation} 
where $\beta \lesssim 1$ is the speed of the jet in units of the speed of light, which is connected to the Lorentz factor through, 
\begin{equation}
\Gamma=\frac{1}{\sqrt{1-\beta^2}}.
\end{equation}
The distribution obtained from our optimal model is plotted, in each case, with the thick black solid line. To give a sense of the uncertainty in these distributions due to the uncertainty in our model parameters we plot, with other line types and colors, the resulting distributions when each model parameter is at the limit that still gives apparent jet speed and redshift distributions acceptable within our 5\% threshold, while all other parameters are kept at their optimal value. 
 
\begin{table}
\setlength{\tabcolsep}{11pt}
\centering
  \caption{ Limits of the model parameters presented as deviations from the optimal value, for which the 5\% K-S test requirement is still met. }
  \label{tab:Parameter limits}
\begin{tabular}{@{}cccc@{}}
 \hline
      & BL Lacs &  FSRQs  \\
  \hline
     $a_{min}$ &  0.738-1.46  & 0.57-0.50 \\
$a_{max}$  & 0.738+0.41  & 0.57+0.12 \\
 $A_{min}$ & 2.251-0.78  & 2.6-0.245  \\
 $A_{max}$ & 2.251+0.68  & 2.6+0.185 \\
  $\tau_{min}$ & - & 0.26-0.003 \\
  $\tau_{max}$ & - & 0.26+0.068 \\
\hline
\end{tabular}
\end{table}

\section{Single-Blazar Doppler Factor Estimates}\label{DFest}

In this section, we review various techniques that have been used in the literature to derive Doppler factor estimates for individual blazars. 

\subsection{Inverse Compton Doppler Factors}

The inverse Compton Doppler Factor $\delta_{IC}$ \citep{Ghisellini1993} is derived based on the requirement that the Synchrotron self-Compton (SSC) flux density should not exceed the observed flux density at high frequencies. The SSC emission consists of photons produced by inverse-Compton upscattering of synchrotron photons by the same relativistic electrons that produce them, and, in that sense, is a guaranteed high-energy component in any region containing magnetic fields and relativistic electrons. 

Assuming a power law energy distribution for the electrons, homogeneous magnetic fields, and the observation frequency ($\nu_m$) to be the self-absorption frequency of the core component dominating at that frequency, the Doppler factor would be:
\begin{equation}
\delta_{IC}=f(\alpha)F_m\left[\frac{ln(\nu_b/\nu_m)}{F_\chi\theta_d^{6+4\alpha}\nu_\chi^\alpha\nu_m^{5+3a}}\right]^{1/(4+2\alpha)}(1+z),
\label{delta_ic}
 \end{equation} 
where $F_m$ is the synchrotron flux density at $\nu_m$ and $F_\chi$ the X-ray flux density both in Jy, $\theta_d$ the angular size of the core in milli arcseconds, $\nu_\chi$ is in keV,  $\nu_m$ is in GHz and $\nu_b$ is the synchrotron high energy cutoff which is assumed to be $10^{14}$ Hz. The function f($\alpha$) is given by \citep{Ghisellini1987} to be $f(\alpha)\simeq 0.08\alpha+0.14$.

Equation (\ref{delta_ic}) is applicable in the case of a discrete jet ($p=3+\alpha$). For the continuous jet case ($p=2+\alpha$) the Doppler factor is related to the one of (Eq.\ref{delta_ic}) by
\begin{equation}
\delta_{2+\alpha}=\delta_{3+\alpha}^{(4+2\alpha)/(3+2\alpha)}.
\label{delta_ic_cont}
\end{equation}
A more detailed description of the model can be found in \citet{Ghisellini1993,Guijosa1996}.

\subsection{Equipartition Doppler Factors}

Equipartition Doppler factors \citep{Readhead1994} use the assumption of equipartition between electrons and magnetic fields in a radio emission region to calculate an intrinsic brightness temperature and, from there, a Doppler factor through comparison to an actual observed brightness tempearature. Different incarnations of this method differ in the way the angular size of the emission is calculated region (direct observation through VLBI or variability timescales and causality arguments.)

\subsubsection{VLBI Equipartition Doppler Factors}

The angular size of a uniform self-absorbed source in order to have equipartition of the radiating particles and the magnetic field, or else the equipartition angular size is \citep{Scott1977},
\begin{eqnarray}
\theta_{eq}&=&10^3(2h)^{1/17}F(\alpha)[1-(1+z)^{-1/2}]^{-1/17}\nonumber\\
&{\times}&S_p^{8/17}(1+z)^{(15-2\alpha)/34}(\nu_p\times 10^3)^{-(2\alpha+35)/34} {\rm mas}, \nonumber\\
\label{theta_eq}
\end{eqnarray}
where $h$ is the dimensionless Hubble parameter, $S_p$ is in Jy and $\nu_p$ in GHz. The function $F(\alpha)$ is described in \citet{Scott1977}. $S_p$ and $\nu_p$ have not been corrected for the beaming effect, thus the observed values are related to the intrinsic ones through $S_p=\delta^{-3}S_{obs}$ and $\nu_p=\delta^{-1}\nu_{obs}$. Assuming the observed angular size is $\theta_d=\theta_{eq}$ the equipartition Doppler factor is,
\begin{eqnarray}
\delta_{eq}&=&\left\{\left[10^3F(\alpha)\right]^{34}([1-(1+z)^{-1/2}]/2h)^{-2}(1+z)^{15-2\alpha}\right.\nonumber\\
&{\times}&\left. S_{obs}^{-16}\theta_d^{-34}(v_{obs}\times 10^3)^{-(2\alpha+35)} \right\}^{1/(13-2\alpha)}.
\label{doppler_eq}
\end{eqnarray}
The equipartition Doppler factor can also be expressed as the ratio of the observed brightness temperature over the maximum intrinsic brightness temperature ($T_{b,int}$). Since for powerful synchrotron radio sources $T_{b,int}$ is equal to the equipartition temperature ($T_{eq}$):
\begin{equation}
\delta_{eq}=\frac{T_{b,obs}}{T_{eq}}\,.
\label{equip_equation}
\end{equation}
A detailed description of the method can be found in \citet{Readhead1994,Guijosa1996,Britzen2007}.

\subsubsection{Variability Doppler Factors}

In this case, the time evolution of a radio flare is used to calculate the brightness temperature of the emission region. The Doppler factor is then obtained by setting that variability brightness temperature equal to the equipartition brightness temperature. Detailed descriptions of this technique \citet{Valtaoja1999,Lahteenmaki1999-II,Lahteenmaki1999-III,Hovatta2009}. For their analysis,  they use long flux density curves, decomposing them to exponential flares of the form,
\begin{equation}
\Delta S(t)=\left\{
\begin{tabular}{lr}
$\Delta S_{max}e^{(t-t_{max})/\tau}$, & $t<t_{max}$\\
$\Delta S_{max}e^{(t_{max}-t)/1.3\tau}$,&$ t>t_{max}$
\end{tabular}\right.
\end{equation}
where $\Delta S_{max}$ is the maximum amplitude of the flare, $t_{max}$ is the epoch of the flare maximum and $\tau$ is the rise time of the flare defined as $\tau=dt/d(\ln{S})$ in days. The observed variability brightness temperature of the source $T_{b,var}$ is,
\begin{equation}
T_{b,var}=1.548 {\, \rm K} \times 10^{-32}\frac{\Delta S_{max}d_L^2}{\nu^2\tau^2(1+z)},
\end{equation}
where $\Delta S_{max}$ is in Jy,  $\nu$ is the observed frequency in GHz, and $d_L$ is the luminosity distance in meters. Using $T_{b,int}=5\times 10^{10}$ K \citep{Readhead1994}, they calculate the variability Doppler factor from 
\begin{equation}
\delta_{var}=\left(\frac{T_{b,var}}{T_{b,int}}\right)^{1/3}.
\label{Tvdoppler}
\end{equation}

\subsection{Single-component Causality Doppler Factors}

The underlying assumption of this method is that the variability timescale of a resolved jet component is determined by the light travel time across the component, rather than loss processes \citet{Jorstad2005,Jorstad2006}.  This method relies upon the observational determination of both the angular size of the component and the variability timescale, defined as:
\begin{equation}
\Delta{t}_{var}=\frac{dt}{ln(S_{max}/S_{min})},
\end{equation}
where $S_{max}$ and $S_{min}$ are the measured maximum and minimum flux densities, respectively, and dt is the time between $S_{max}$ and $S_{min}$. The Doppler factor can be calculated from:
\begin{equation}
\delta_{var}=\frac{sd_{L}}{c\Delta{t}_{var}(1+z)},
\end{equation}
where $d_{L}$ is the luminosity distance, and s is the angular
size of the component, equal to $1.6a$ for $a$ Gaussian, equal to the full width at half maximum, measured at the epoch of $S_{max}$. After calculating a Doppler factor for each observed component for a specific source, the weighted average of these values is assigned to a source, with the weights being inversely proportional to the uncertainty in apparent velocity of each component.
 
This method is resource-expensive in that it requires multiepoch VLBI monitoring for each source. For this reason, Doppler factors at this stage have been calculated by this method, to our knowledge, only for 5 BL Lac objects and 8 FSRQs. As a result, a statistical evaluation for this method is rendered impractical due to low statistics. 

\subsection{Gamma-ray Opacity Doppler factors}

The calculation of $\gamma$-ray opacity Doppler factors is based on the requirement that the $\gamma-$ray emission region must be transparent to gamma rays \citep{Mattox1993,Fan2013,Fan2014}. The process responsible for $\gamma$-ray absorption is pair production due to the interaction between $\gamma$-ray photons and $X$-ray photons. They are a lower limit to the true Doppler factors. This calculation involves the assumptions that the emission region is spherical and that X-rays are produced in the same region as gamma rays. Causality arguments are used to connect variability timescales with the emission region size. The Doppler factor is then given by 
\begin{eqnarray}
\delta_{\gamma}&{\geq}&\left[1.54\times{10^{-3}}(1+z)^{4+2\alpha}\left(\frac{d_L}{ \rm Mpc}\right)^2  \right]^{1/(4+2\alpha)}\nonumber\\ 
&{\times}&\left[\left(\frac{\Delta{T}}{\rm hours }\right)^{-1}\left(\frac{F_{keV}}{ \rm \mu Jy }\right)\left(\frac{E_{\gamma}}{\rm Gev}\right)^{\alpha}\right]^{1/(4+2\alpha)},
\end{eqnarray}
where $\alpha$ is the X-ray spectral index, $F_{keV}$ the flux density at 1 keV in $\mu$Jy, $E_{\gamma}$ is the energy at which the $\gamma$-rays are detected in GeV, $d_L$ the luminosity distance as described in \S \ref{DFdistr}, and $\Delta{T}$ is the time scale in hours defined as:
\begin{equation}
\Delta{T}=\frac{(1+z)R}{c\delta_{\gamma}},
\end{equation}
where R is the size of the emission region. A variation of this method uses UV photons as the target photon field and similar arguments to derive a Doppler factor. 

In this work, we do not statistically test this technique, because the calculation of a statistical Doppler factor distribution requires a well-defined, 15 GHz radio flux-limited sample\footnote{Because for these sources the spectral index is very close to zero we have used $F_\nu{\approx}F_{15\rm GHz}$ for nearby frequencies.}. Because of the significant scatter in the radio/gamma-ray flux correlation \citep{Pavlidouff}, it is not straight-forward to calculate a single radio flux limit for a gamma-ray selected sample, even if the latter is flux-limited. At the same time, the fact that gamma-ray opacity arguments can only provide lower limits to the true Doppler factor complicates the statistical comparison of these Doppler factors to other datasets and models. However, such a comparison would in principle be very interesting, especially if it could confirm whether  $\gamma$-ray and radio emitting regions have different outflow velocities \citep{Georganopoulos2003-I,Georganopoulos2003-II,Georganopoulos2003-III,Georganopoulos2003-IV,Giannios2009} and hence different Doppler factors.

\section{Comparison with Statistical Doppler Factors}\label{comparison}

In this section we compare our results on Doppler factor distributions from our blazar population models with data on Doppler factor estimates through different techniques found in the literature. Every data set we have tested was separated into two populations, one for the BL Lac objects and one for FSRQs.  It was shown in Paper I that the model can adequately describe blazars as a population; however  all derived distributions are sample-specific and flux-limit dependent. For this reason, if we want to compare any derived distribution with data, including the Doppler factor distributions,  the flux-limit of the sample at hand must be taken into account. As a result, for each dataset and each object class, a  distribution of statistical Doppler factors was derived by using our model and the flux limit of the corresponding sample of the data set we are comparing with. All the data we compare with are from flux-limited samples, either 0.35 Jy, 1 Jy or 2 Jy. Comparisons between statistical and single-blazar estimated Doppler factor distributions are made with the use of the Kolmogorov-Smirnov (K-S) test. All the values presented in this work represent the probability of the single-blazar estimates having been drawn from the corresponding statistical Doppler factor distribution. 
In order to accept the consistency statement we require the probability value to be higher than 5\%. 

\begin{figure}
\resizebox{\hsize}{!}{\includegraphics[scale=1]{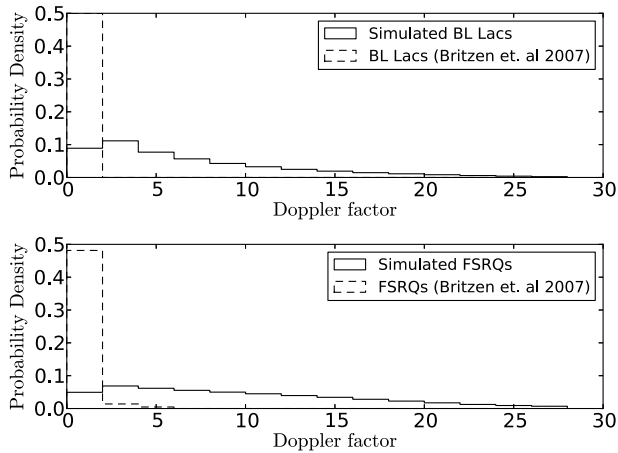} }
\caption{Probability density functions of statistical Doppler factors (solid line) and equipartition Doppler factors \citep{Britzen2007} (dashed line) for the BL Lac (upper panel) and the FSRQ (lower panel) sample.}
\label{pdf_britzen_doppler_eq}
\end{figure}

\begin{figure}
\resizebox{\hsize}{!}{\includegraphics[scale=1]{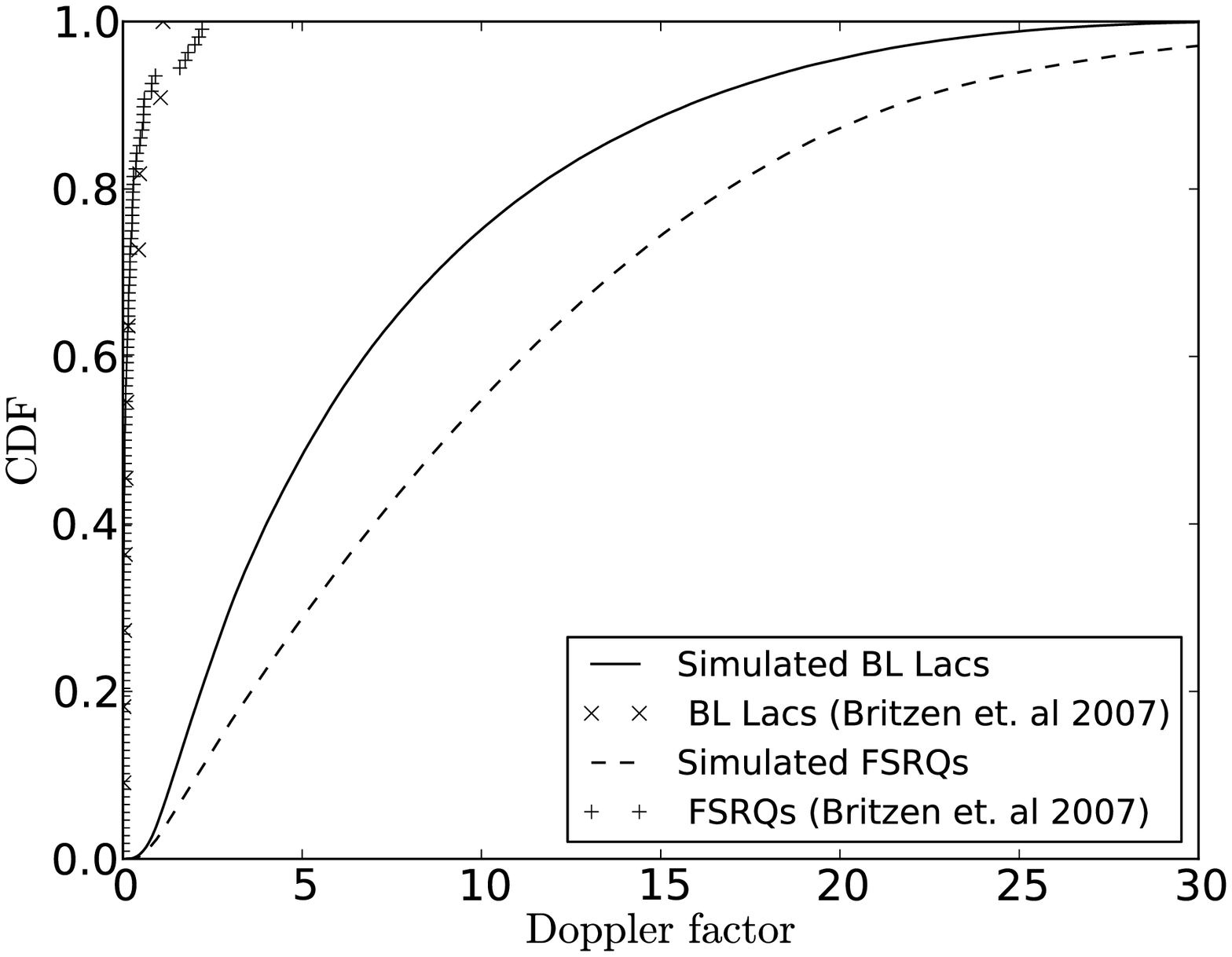} }
\caption{Cumulative distribution functions of statistical Doppler factors for BL Lacs (solid line) and FSRQs (dashed line), overplotted with  equipartition Doppler factors \citep{Britzen2007} (X for BL Lacs, + for FSRQs). }
\label{britzen_doppler_eq}
\end{figure}

\begin{figure}
\resizebox{\hsize}{!}{\includegraphics[scale=1]{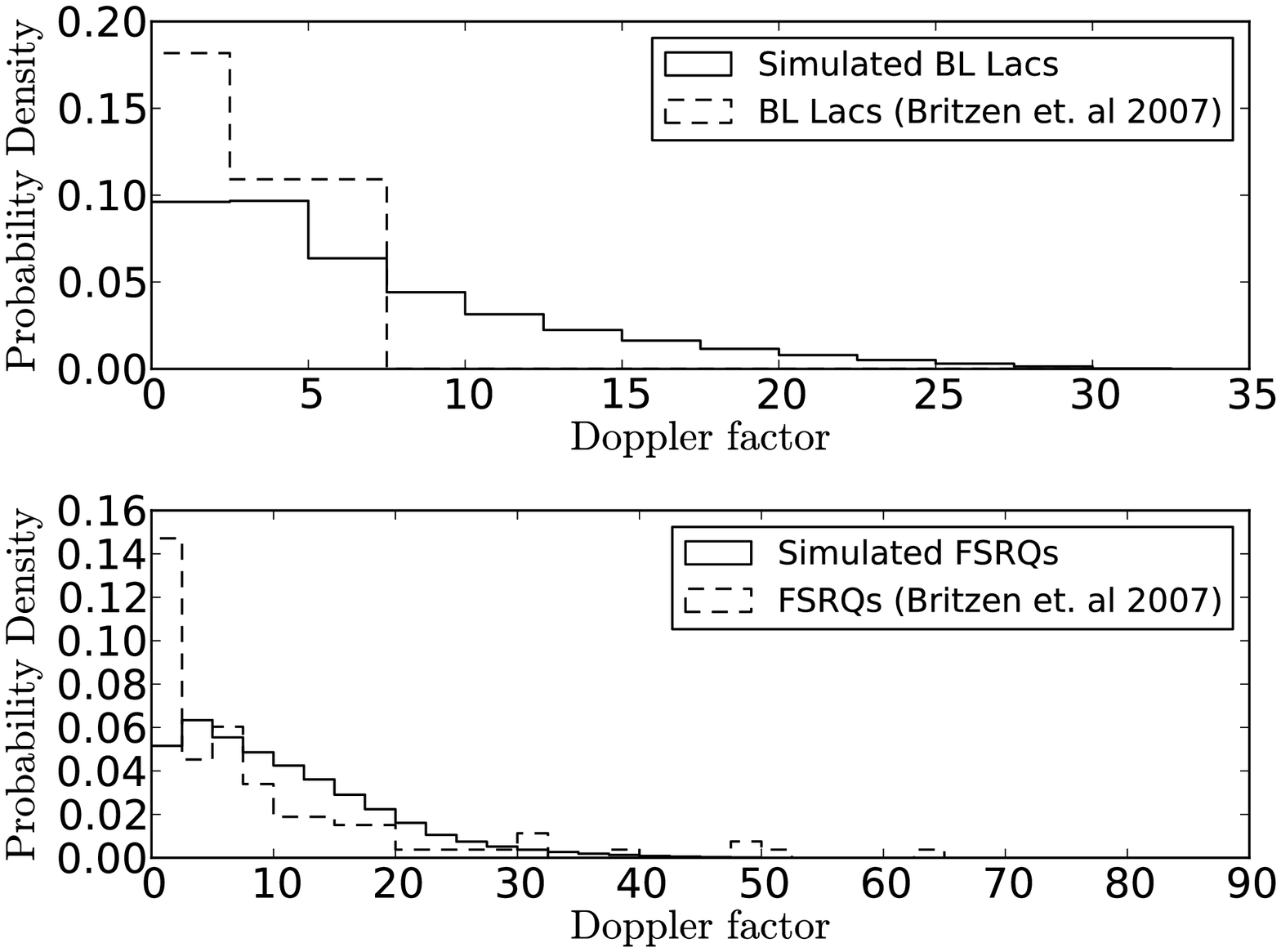} }
\caption{Probability density functions of statistical Doppler factors (solid line) and inverse Compton Doppler factors \citep{Britzen2007} (dashed line) for the BL Lac (upper panel) and the FSRQ (lower panel) sample.}
\label{pdf_britzen_doppler}
\end{figure}

\begin{figure}
\resizebox{\hsize}{!}{\includegraphics[scale=1]{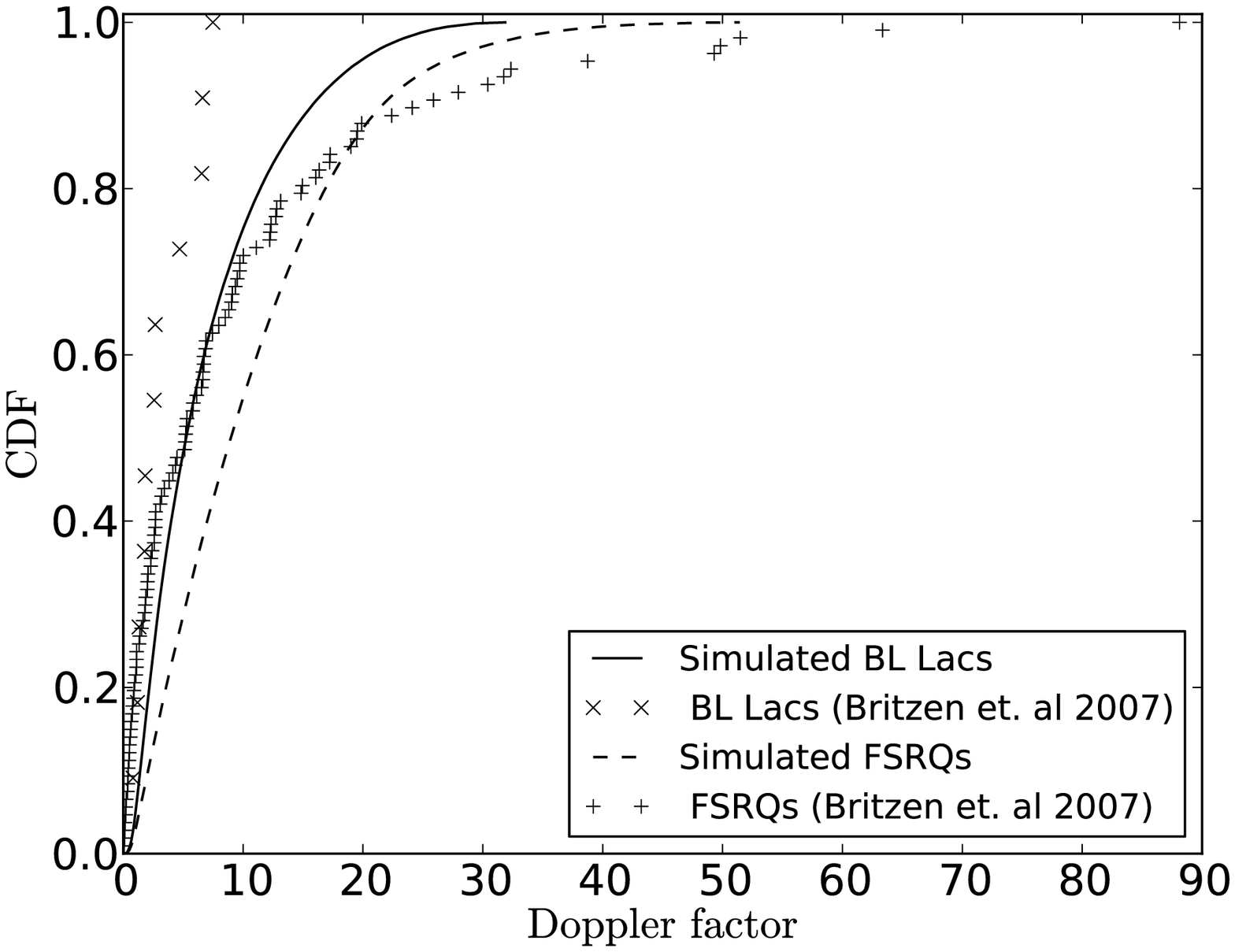} }
\caption{Cumulative distribution functions of statistical Doppler factors for BL Lacs (solid line) and FSRQs (dashed line), overplotted with  inverse Compton Doppler factors \citep{Britzen2007} (X for BL Lacs, + for FSRQs). }
\label{britzen_doppler}
\end{figure}

\begin{figure}
\resizebox{\hsize}{!}{\includegraphics[scale=1]{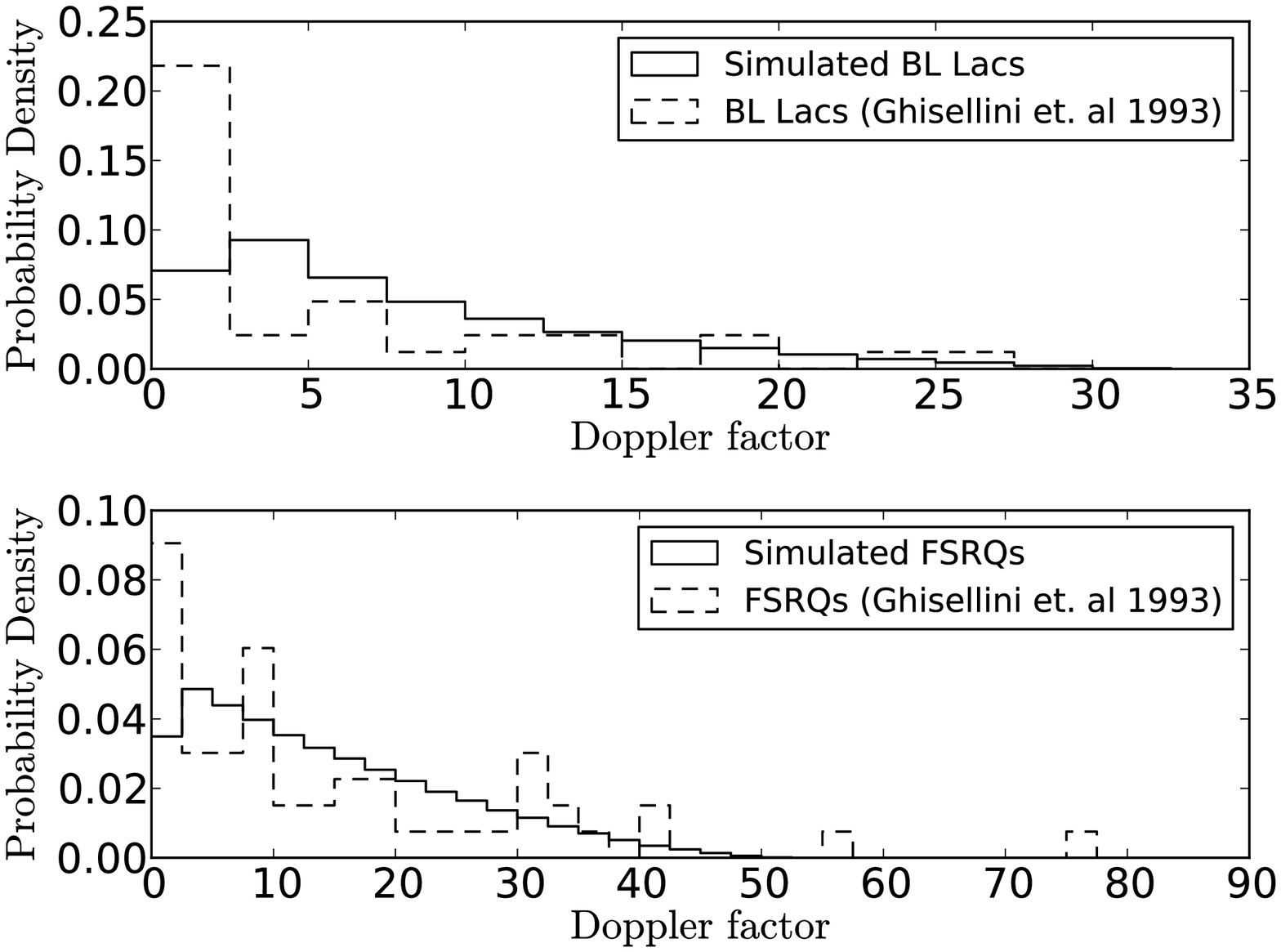} }
\caption{Probability density functions of statistical Doppler factors (solid line) and inverse Compton Doppler factors  \citep{Ghisellini1993} for the BL Lac (upper panel) and the FSRQ (lower panel) sample.
\label{pdf_ghisellini_doppler}}
\end{figure}

\begin{figure}
\resizebox{\hsize}{!}{\includegraphics[scale=1]{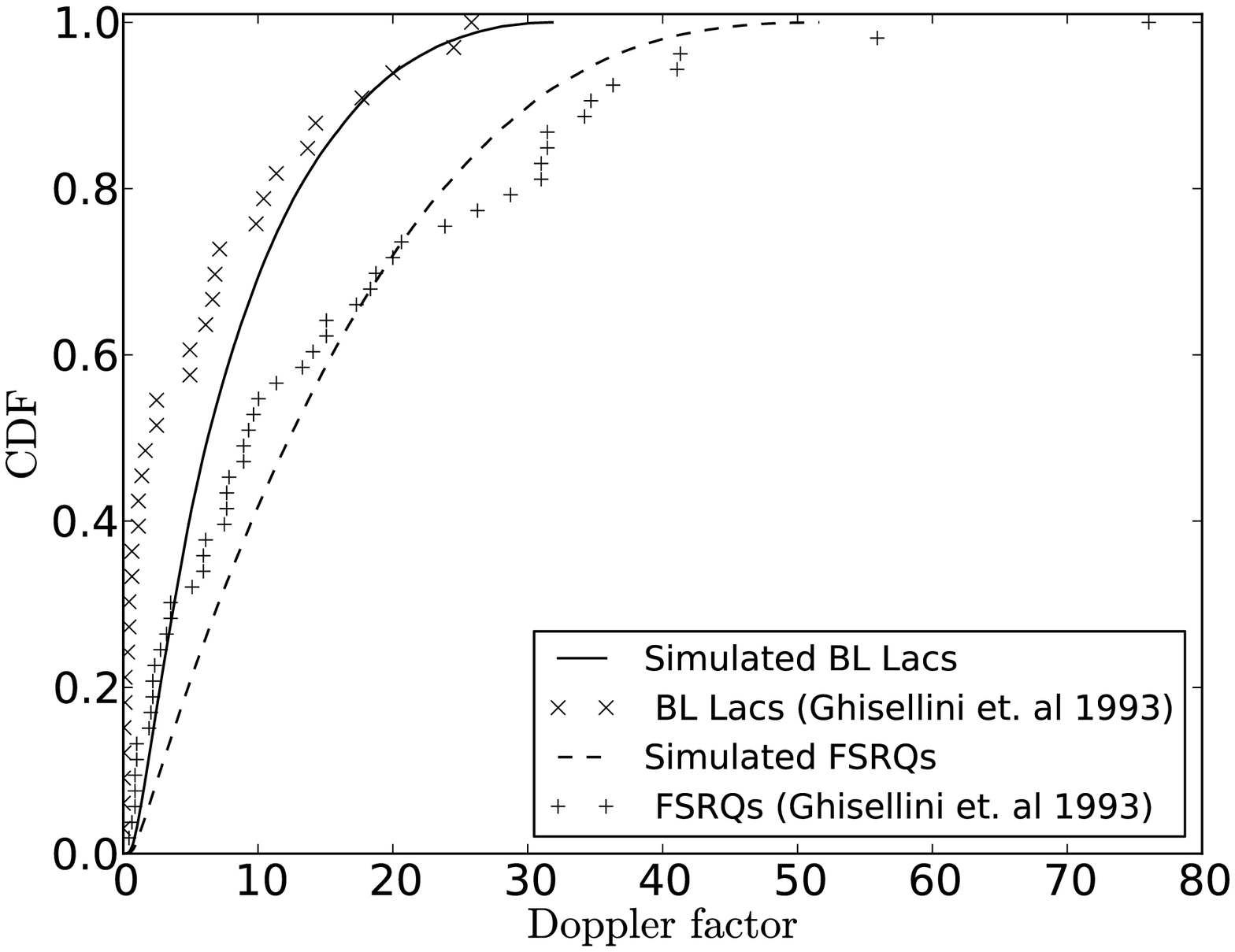}}
\caption{Cumulative distribution function of the statistical Doppler factors for BL Lacs (solid line) and FSRQs (dashed line), overplotted with inverse Compton Doppler factors from \citet{Ghisellini1993} (X for BL Lacs, + for FSRQs). }
\label{ghisellini_doppler}
\end{figure}

\begin{figure}
\resizebox{\hsize}{!}{\includegraphics[scale=1]{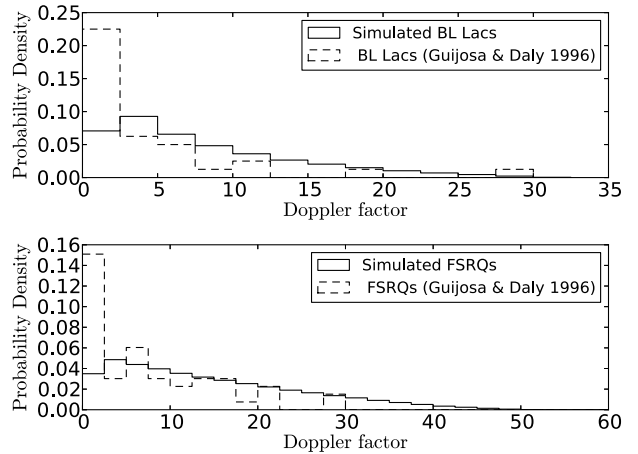}}
\caption{Probability density function of statistical Doppler factors (solid line) and equipartition Doppler factors from \citet{Guijosa1996} (dashed line) for the BL Lac sample (upper panel) the FSRQ sample (lower panel).}
\label{pdf_daly_doppler}
\end{figure}

\begin{figure}
\resizebox{\hsize}{!}{\includegraphics[scale=1]{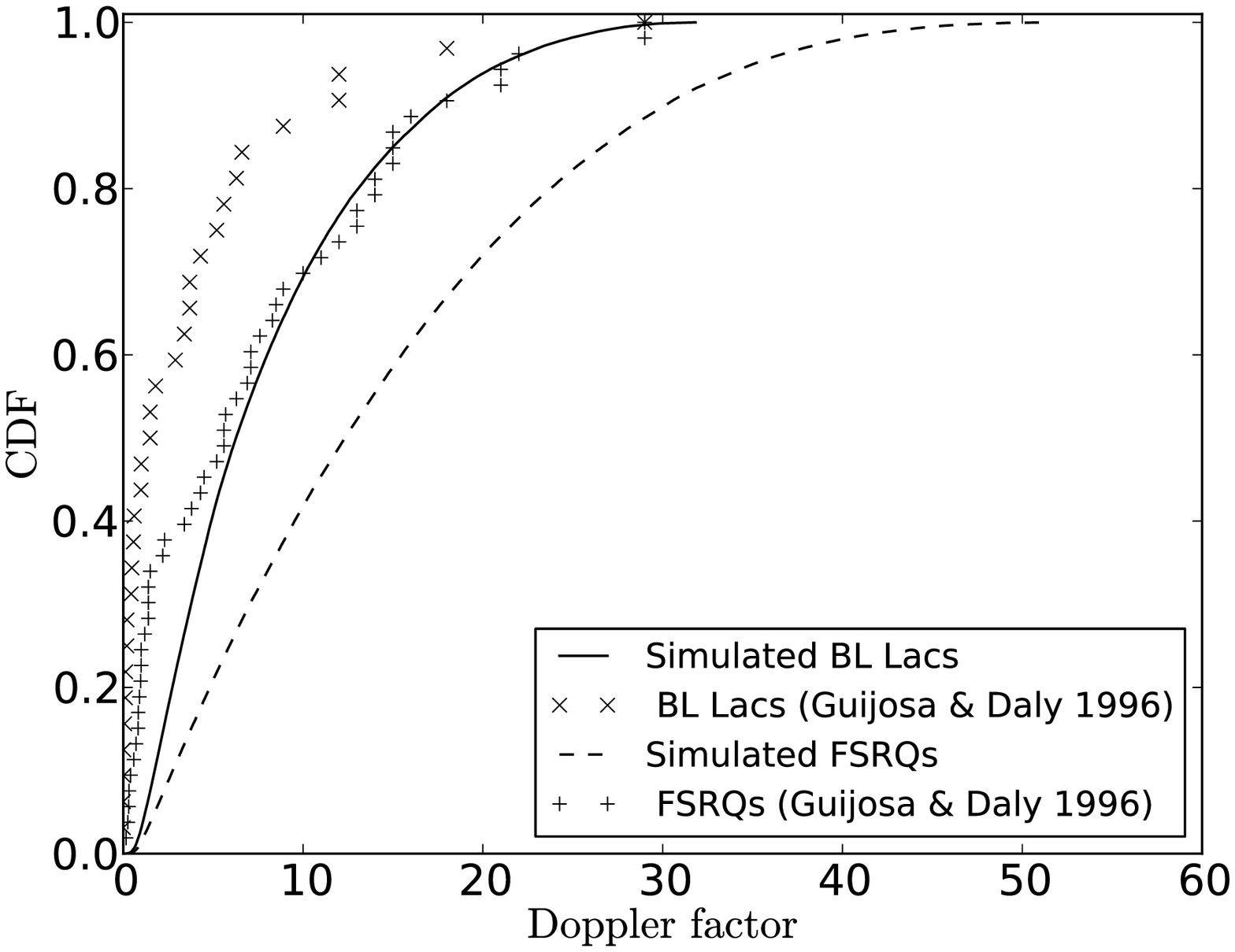} }
\caption{Cumulative distribution function of the statistical Doppler factors for BL Lacs (solid line) and FSRQs (dashed line) overplotted with equipartition Doppler factors from \citet{Guijosa1996}  (X for BL Lacs, + for FSRQs). }
\label{daly_doppler}
\end{figure}

We start by comparing our results with the data given in \citet{Britzen2007}. They provide data for both inverse Compton ($\delta_{IC}$) and equipartition ($\delta_{eq}$) Doppler factors. The sample consists of 11 BL Lac objects and 108 FSRQs. The flux-limit for this sample is 0.35 Jy \citep{Taylor1996}. 

The data set for the $\delta_{eq}$ features extremely low values, with the highest value for the BL Lacs $\delta_{eq}\sim 1.11$ and for the FSRQs $\delta_{eq}\sim 4.72$. A K-S test confirmed that these distributions are not consistent with what is expected for these populations, with the p-value for the BL Lacs $\sim 1.22\times 10^{-7}\%$ and for the FSRQs $\sim 1.16\times 10^{-78}\%$. For the case of the $\delta_{IC}$ we have excluded source 1732+389 from the FSRQ sample for being an outlier ($\delta_{IC}\approx 276$). The consistency between estimated and statistical Doppler factor distributions is rejected for FSRQs, with a K-S test returning a p-value of $\sim 2.15\times 10^{-5}\%$.  The value for the K-S test for the BL Lac sample is $\sim 6.7\%$. Note however that the statistics in the BL Lac sample are extremely low. 

We plot the probability density function (PDF) and the cumulative distribution function (CDF) for the $\delta_{eq}$ case, in Figs.~\ref{pdf_britzen_doppler_eq} and \ref{britzen_doppler_eq} and for the $\delta_{IC}$ case in Figs.~\ref{pdf_britzen_doppler} and \ref{britzen_doppler} . 

We proceed to the data set analysed by \citet{Ghisellini1993} for the inverse Compton Doppler factor. The sample consists of 33 BL Lac objects and 53 FSRQs. The flux-limit for this sample is 1 Jy \citep{Kuehr1981}. We corrected for the continuous jet case (which we also assume in our statistical model) using eq. \ref{delta_ic_cont}. The KS-test returns a p-value of $\sim 4\times 10^{-3}\%$ for the BL Lacs and $\sim 10.36\%$  for the FSRQs. We plot the PDF and CDF for the two classes in Figs.~\ref{pdf_ghisellini_doppler} and \ref{ghisellini_doppler}.

We next test equipartition Doppler factors calculated by \cite{Guijosa1996}. The sample consists of 32 BL Lac objects and 53 FSRQs. The flux-limit for this sample is 1 Jy \citep{Ghisellini1993}. The value of the KS-test for the BL Lac objects is $\sim 1.5\times 10^{-4}\%$ while for the FSRQs $\sim 7.6\times 10^{-3}\%$. Figures \ref{pdf_daly_doppler} and \ref{daly_doppler} show the simulated and estimated probability density (fig. \ref{pdf_daly_doppler}) and cumulative distribution (fig. \ref{daly_doppler}) functions.

\begin{figure}
\resizebox{\hsize}{!}{\includegraphics[scale=1]{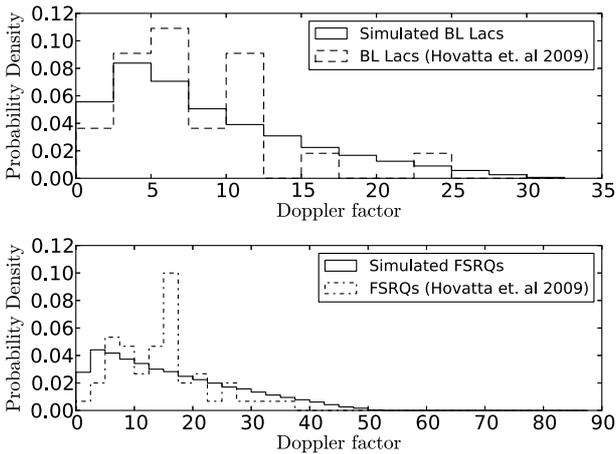}}
\caption{Probability density function of statistical Doppler factors (solid line) overplotted with variability Doppler factors using equipartition from \citet{Hovatta2009} (dashed) for the BL Lac sample (upper panel) and the FSRQ sample (lower panel).}
\label{pdf_hovatta_doppler}
\end{figure}

\begin{figure}
\resizebox{\hsize}{!}{\includegraphics[scale=1]{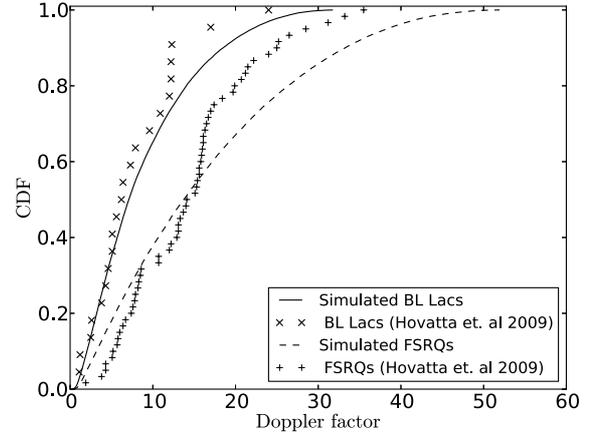}}
\caption{Cumulative distribution function of statistical Doppler factors for the BL Lacs (solid line) and FSRQs (dashed line) overplotted with variability Doppler factors using equipartition from \citet{Hovatta2009}  (X for BL Lacs, + for FSRQs).}
\label{hovatta_doppler}
\end{figure}

Finally, we test variability Doppler factors (using the equipartition brightness temperature to derive a Doppler factor) using data from \cite{Hovatta2009}. They use a  sample consisting of 22 BL Lac objects and 60 FSRQs. The flux-limit for this sample is 2 Jy \citep{Valtaoja1992}. For the BL Lac population, the value of the K-S test is  $\sim 54\%$ while for the FSRQ $\sim 14\%$. The PDFs and CDFs for the comparison are shown in Figs.~\ref{pdf_hovatta_doppler} and \ref{hovatta_doppler} respectively. This is the method that gives the best agreement with the statistical Doppler factors. 

\section{Discussion}\label{discussion}

We have compared, in \S\ref{comparison},  results from our statistical, population model for blazars, with data provided in the literature for three distinct methods of calculating Doppler factors. The results of this comparison show that one of the methods, variability Doppler factors $\delta_{var}$, is consistent with statistical Doppler factors (K-S test BL Lac $\sim 54\%$, FSRQs $\sim 14\%$) whereas the other two methods seem to be drawn from completely different distributions (K-S test for both populations $\sim\leq 10^{-3}\%$). This result is in agreement with \cite{Lahteenmaki1999-III} arguing that variability Doppler factors using equipartition are a better and more accurate estimate of the Doppler factors of beamed sources than equipartition Doppler factors that rely on direct measurement of the angular size of the emission region, due to a weaker dependence on the observed brightness temperature (third root compared to first power). 

Of all other Doppler factor estimation techniques we tested, there were only two cases where our statistical analysis did not indicate an inconsistency between our statistical model and the data: the BL Lac sample in \citet{Britzen2007} and the FSRQ sample in \citet{Ghisellini1993}.

In the first case, even though the value of the K-S test is marginally acceptable ($\sim 6\%$), it is clear from the low value of the Doppler factors (maximum value $\sim 7$) and from Figs.~\ref{pdf_britzen_doppler} and \ref{britzen_doppler} that the agreement is far from good; however, the low number of sources in the BL Lac sample (11 sources) is preventing a strong conclusion either way through the K-S test.

 In the second case, the test gives a probability of $\sim$10\% for consistency, which is above the limit set in \S\ref{comparison} and thus considered acceptable. It is interesting to note that the results of the comparison are very different between the two population classes (BL Lacs  and FSRQs). Agreement of their results with our optimal model for the BL Lac sample is rejected at the $\sim 4\times 10^{-3}\%$ level. As discussed by \cite{Britzen2007}, $\delta_{IC}$ is equal to the real Doppler factor only if all of the observed flux in X-rays is produced through the SSC process. In any other case, $\delta_{IC}$ represents only a lower limit. If all other assumptions entering the $\delta_{IC}$ calculation hold, our findings would suggest that part of the X-ray flux is produced through other mechanisms for many sources in the BL Lac sample. This is consistent with our general understanding of these classes of sources: FSRQs are low synchrotron peaked (LSP) sources \citep{Abdo2010-II} so it is reasonable to expect the largest fraction of their X-ray emission to be of inverse-Compton origin; on the other hand, many  BL Lacs are intermediate synchrotron peaked (ISP) and high synchrotron peaked (HSP), so part of their X-ray flux can be produced by synchrotron emission, which would result in the inverse Compton Doppler factors underestimating the true Doppler factor of their jets.
 Indeed, it is clear from Fig.~\ref{ghisellini_doppler} that the Doppler factors of the BL Lacs are underestimated compared to the expectations from our optimal population model. 

It is troubling that the consistency of $\delta_{IC}$ with the expectations from our population model for FSRQs is so different between the \cite{Ghisellini1993} sample and the \cite{Britzen2007} sample. While the first is acceptable, the second is rejected with a probability of $\sim 2.15\times 10^{-5}\%$. Six orders of magnitude in difference might indicate poor sample selection or/and large errors in measurements. This difference might also be due to the time difference between radio and X-ray flux measurements. The main assumption of the inverse Compton Doppler factor method is that synchrotron-emitting electrons up-scatter synchrotron photons to X-rays. The X-ray and radio flux densities must therefore be measured at the same time. Any time difference in observations might result in non-corresponding flux densities. A better evaluation of this technique could be achieved by systematically pursuing simultaneous radio and X-ray observations for a radio-selected flux-limited sample of low-synchrotron peaked sources.

For those Doppler techniques and samples that the hypothesis of being drawn from our optimal statistical model is rejected by the K-S test, we have also performed comparisons with distributions drawn from a statistical model with parameters other than the optimal, but still within the limits of acceptability discussed \S \ref{DFdistr}.  Since in these cases there is an excess of low values for the Doppler factors, we have compared them with distributions resulting from the model using $a_{max}$ and the model using $A_{min}$ since these give a higher fraction of low Doppler factor values than our optimal model (see Figs.~\ref{bllim} and \ref{qsolim}) for both populations. A K-S test indicated that agreement with these distributions is also rejected.

\subsection{Error analysis}
\begin{figure}
\resizebox{\hsize}{!}{\includegraphics[scale=1]{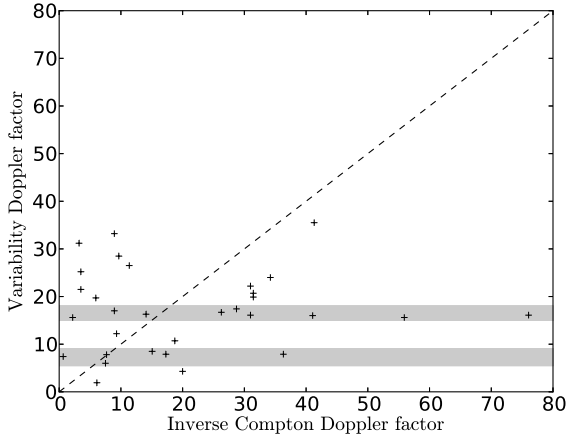} }
\caption{Variability Doppler factors plotted against inverse Compton Doppler factors for the FSRQs that are common between the \citet{Hovatta2009}  and \citet{Ghisellini1993} samples. The gray shaded areas indicated the regions of pileups.}
\label{ghis_hov}
\end{figure}

Although the variability Doppler factors using equipartition result in the best agreement with our statistical model (and, by extension, with MOJAVE results on jet speeds), the agreement is not perfect. First, there is a pileup of sources between Doppler factors of 10 and 16 in both classes of sources. In the FSRQ case, there seems to be another pileup between 5 and 8 and a deficit of low Doppler factor values (see Fig.~\ref{hovatta_doppler}). Such pileups appear to be systematics-related. For example: high Doppler factors result in very compressed timescales, and the fastest flare in a source may not be resolvable at a given cadence. Higher time resolution in the flux density curves will provide more accurate results.

Similarly, for the other case where we have a Doppler factor estimation method that is not inconsistent with our optimal model (the inverse Compton Doppler factors of \citealp{Ghisellini1993} for FSRQs), the agreement is also not perfect. There is a concentration of sources at low Doppler factor values compared to the expectation from our statistical model. This is expected if in the sample under study there is a considerable number of sources where the SSC emission is not the sole source of X-rays; or, if there are errors entering due to the non-simultaneity of X-ray and radio measurements. 

In Fig.~\ref{ghis_hov} we plot variability Doppler factors using equipartition from \citet{Hovatta2009} and inverse Compton Doppler factors from \citet{Ghisellini1993} for the 31 common FSRQ sources in their samples. We have indicated the regions of pileup of sources discussed above for the variability Doppler factor method in grey. Although a correlation can be seen for the remaining sources, the statistics are very low for a strong statement to be made. We note however that such pileups in the distribution of Doppler factors in a flux-limited sample where none are expected from a population model can be  a good indicator regarding the sources for which the results of a particular technique should be used with caution. 

 As discussed in \S \ref{DFdistr}, neither the Doppler factor estimates discussed here, nor the assumptions on which they have been based have been used in any way in the optimization of the population models we used to produce simulated Doppler factor distributions. For this reason, the simulated and estimated Doppler factor distributions are statistically independent. This provides us with a unique opportunity: we can use the observed and simulated (assumed to be ``intrinsic'') Doppler factor distributions to derive the (average) error on the Doppler factor estimates. This is especially important since errors on the Doppler factor estimates are difficult to calculate and are not provided in the original analyses from which the Doppler factor estimates used in this work are taken. We next provide such an error analysis for the two methods (variability and IC Doppler factors) that are consistent at the population level with our simulated distributions.

We construct the cumulative distribution function of the model Doppler factors, for each population. We set a constant fractional error p (from 0 to 1) for each Doppler factor estimate that is common for all values in each method. For the case of normal errors we draw a random value from the model Doppler factor CDF, which serves as the ``true'' Doppler factor of a source and the mean of the error distribution. The standard deviation, or shape parameter of each distribution is the mean multiplied by the fractional constant (p). Then we drew a random value from that distribution. Any negative values were rejected, so, strictly, the error distribution in this case is a truncated Gaussian. By following the same procedure for different simulated sources, we created a simulated with-errors sample, which we compared with the corresponding method using the K-S test. We repeated this process with step 0.01 in p in order to assure the fractional constant's parameter space (from 0 to 1) was adequately scanned for different distributions. The distribution which increased the agreement between the simulated with-errors and observed samples the most is the error distribution of each estimate with error the ``best-fit" percentage error which is a function of p. We have also experimented with many other error distributions (log-normal, uniform, width parameters other than p$\delta$ etc.) and tested again whether agreement with the distribution of estimated Doppler factors improves.

\begin{figure}
\resizebox{\hsize}{!}{\includegraphics[scale=1]{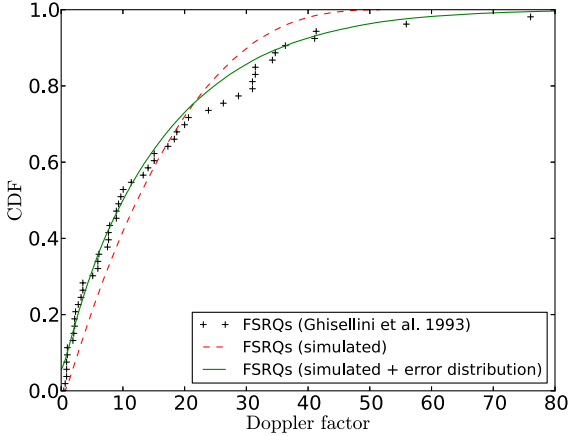}  }
\caption{Cumulative distribution function of the simulated ``errored" sample (solid green) overplotted with the initial simulated sample (dashed red) and the variability Doppler factors \citep{Ghisellini1993} (``+") for the FSRQs.}
\label{ghis_error_qs}
\end{figure}

In the case of the inverse Compton Doppler factor we treated only the FSRQs, since it is the only population that the method can adequately describe.  We found that the error distribution that best describes the method is a normal distribution with mean $\mu=\delta_{IC}$ and standard deviation $\sigma=p\delta_{IC}$ with ``best-fit" fractional constant $p\approx 0.63$, which in this case corresponds to a percentage error of $\sim 63\%$. The K-S test gave a probability value of $94.6\%$ of consistency between the simulated with-errors and  the observed sample. Figure \ref{ghis_error_qs} shows the cumulative distribution function for the simulated with-errors distribution for the FSRQ population.
In \S \ref{comparison}, we found that the probability of consistency between the simulated and observed samples for the $\delta_{IC}$ was $\sim 10.36\%$. Introducing the error distribution, there is a significant rise in the probability of consistency of the two samples. Since the inverse Compton Doppler factor method does not involve fitting and the calculation is performed by inputting parameter values, the error in each estimate depends only on the errors of these values. In this case, standard error propagation methods dictate that the error distribution of each estimate has a Gaussian shape, which is validated by our results. In addition, the error distribution shows that even though the error in each estimate is relatively large ($\sim 63\%$) the overall error is dominated by random errors, which can be attributed to uncertainties in the measurements and/or non-simultaneity of observations. The source of errors in the inverse Compton Doppler factor method will be the subject of a future publication.

For the variability Doppler factors we were unable to find a error distribution able to adequately ``fit'' the data. The reason for this is that the the variability Doppler factor uncertainties appear to be dominated by systematic, rather than random errors.  As discussed above, there are pileups in the distribution of both the BL Lacs and FSRQs (Figs. \ref{hovatta_doppler}, \ref{ghis_hov}), for which there is no satisfactory way of treating in a statistical fashion.

It is obvious, even by eye (Fig. \ref{hovatta_doppler}), that the BL Lac population is less affected by systematic errors. There is a small pileup at $\delta_{var}\sim 12$, but otherwise there seem to be no other prominent features. For this reason we used the BL Lac distribution in order to have a coarse estimate on the random errors of the variability Dopple factors. To achieve this, we constructed the cumulative distribution functions of the variability Doppler factors, and the with-errors distribution, assuming each estimate comes from a normal distribution with mean the value of that estimate ($\mu=\delta_{var}$) and standard deviation the value multiplied by the fractional constant p ($\sigma=p\delta_{var}$). Then, we calculated the distance between the two CDFs (i.e the K-S statistic), but this time, for the derivation of the maximum distance we excluded the three points that lay in the pileup (Fig. \ref{hovatta_doppler}). We found that the shortest distance is achieved with $\sim 30\%$ error in each estimate. Adding error above that increases the distance between  the two CDFs. These results suggest that although the error of the variability Doppler factors is dominated by systematics, the random error in each estimate is approximately  $30\%$. Since there is no difference between BL Lacs and FSRQs in the procedure used to estimate the variability Doppler factor, we expect the FSRQs to have the same random error as the BL Lacs, which will be $\sim 30\%$ as well.

Comparing the two methods discussed above (variability, inverse Compton), we see that the variability Doppler factor estimates have approximately half the percentage  random error of the inverse Compton Doppler factors; making them the most accurate of the two, which is consistent with the analysis in \cite{Lahteenmaki1999-II}. Although the variability Doppler factor method is the most accurate method for describing blazars as a population, the cadence of observations warrants caution in the application of the method and the primary source of systematic errors.

\subsection{Statistical Doppler factors}
We have evaluated whether the theoretically derived Doppler factor distributions can be combined with measurements of the flux density and apparent jet speed of a source to yield an estimate of the Doppler factor in individual sources. 
\begin{figure}
\resizebox{\hsize}{!}{\includegraphics[scale=1]{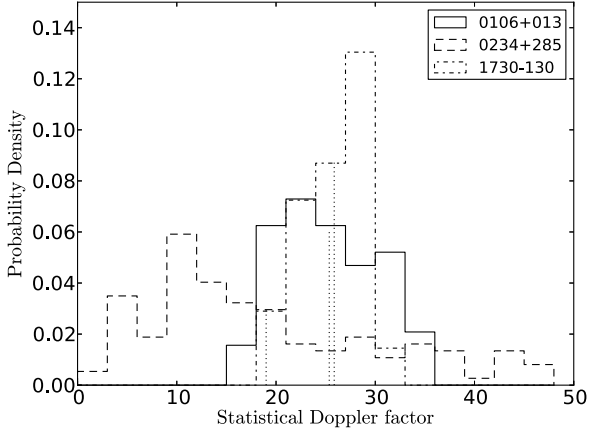}}
\caption{Distribution of Doppler factors for 0106+013 0234+285 and 1730-130. The vertical dotted lines represent the mean values for each of the distributions.}
\label{dist_st_var}
\end{figure}
We created two sub-samples from the common sources in the MOJAVE sample and \citep{Hovatta2009}, for which we also  have flux density measurements \citep{Lister2009}. These consist of 12 BL Lacs and 39 FSRQs respectively. 
For each of these sources, we generated a distribution of Doppler factors as follows: starting from our blazar population model for the relevant class of sources (BL Lacs or FSRQs), we randomly drew luminosities, Lorentz factors, and viewing angles according to their respective distributions. However, instead of keeping only sources that obeyed a specific flux limit, we only kept sources with flux density within $10\%$ of the mean flux density of the source at hand. 

We show these distributions for three of these sources in Fig.~\ref{dist_st_var}. These three sources have the additional property that their variability and inverse Compton Doppler factor estimates are very close to each other. The values of their variability Doppler factors are These sources are 18.4, 16.1, and 10.7, for 0106+013, 0234+285, and 1730-130  respectively. We can see that all three distributions have a very significant spread. The mean and standard deviation of these distributions are $25.36 \pm 4.96$ for 0106+013,  $19.00\pm 11.57$ for 0234+285,  and $25.86\pm 2.88$ for 1730-130.

\begin{figure}
\resizebox{\hsize}{!}{\includegraphics[scale=1]{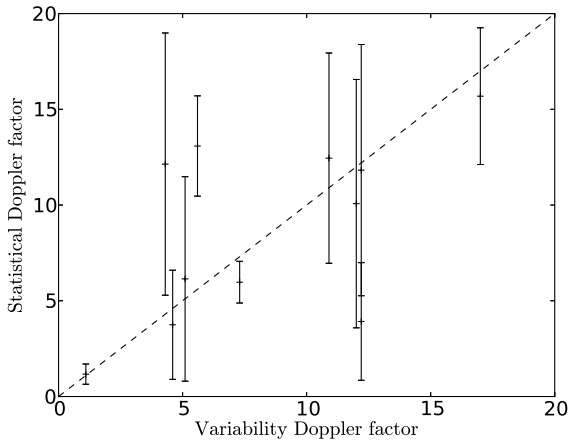}}
\caption{ Statistical versus variability (using equipartition) Doppler factors for the BL Lac sample. The error bars represent 1$\sigma$ of the statistical Doppler factor distribution for each source.}
\label{st_var_bllac}
\end{figure}

\begin{figure}
\resizebox{\hsize}{!}{\includegraphics[scale=1]{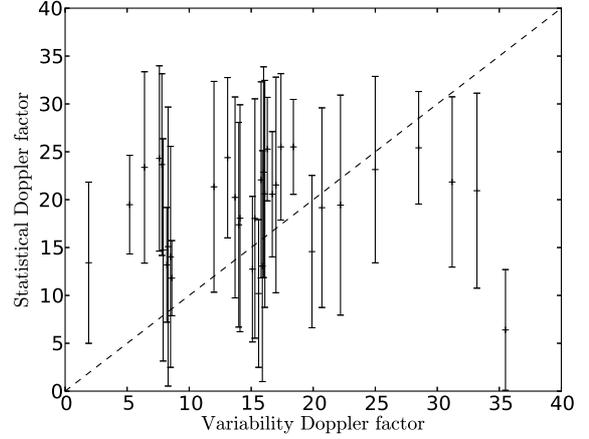}}
\caption{Statistical versus variability (using equipartition) Doppler factors for individual FSRQs The error bars represent 1$\sigma$ of the statistical Doppler factor distribution for each source. }
\label{st_var_qso}
\end{figure}

Figures \ref{st_var_bllac} and \ref{st_var_qso} show the statistical Doppler factors (mean and standard deviation of the Doppler factor distribution produced for each source as described above) plotted against the variability Doppler factors for the BL Lacs and the FSRQs respectively.
We can see that while there is agreement within 2$\sigma$ for most sources, the uncertainties of the statistical Doppler factors are so large that they erase any correlation between the two quantities on a source-by-source basis. We conclude that our population model is not constraining enough when applied to individual sources, and thus we strongly advise against using it to derive information about single objects. 

\section{Conclusions}\label{conclusions}

We have used our population models for BL Lacs and FSRQs to evaluate different techniques of calculating Doppler factors in individual sources. Our conclusions can be summarized as follows. 

\begin{itemize}

\item{Variability Doppler factors using equipartition, when calculated for a flux-limited sample, result in a distribution that is consistent for both samples (FSRQS \& BL Lacs) with the distribution produced by our population model when the same flux limit is applied.  

Since the only observables entering the optimization of our population model are apparent jet speeds and source redshifts, our model contains no assumption regarding variability, causality, or equipartition. The agreement between our model distributions and the distributions of variability Doppler factors points to a self-consistent picture in radio between jet speeds and Doppler factors. Additionally, this agreement can only be achieved if equipartition as discussed by \citet{Scott1977,Readhead1994} indeed holds in blazars as a population. }

\item{Inverse Compton Doppler factors are not inconsistent with the Flat Spectrum Radio Quasars as a population, while they are unable to describe BL Lacs, likely because some of the latter produce a significant fraction of their X-ray flux through synchrotron radiation. This conclusion is also in support of the self-synchrotron Compton model, a main assumption of the inverse Compton Doppler factor method,  being responsible for most of the X-ray flux in a large fraction of FSRQs.}

\item{Exploring the error distribution of the methods that can adequately describe blazars as a population, we found that: for the inverse Compton Doppler factors, the error distribution  is a (truncated) normal distribution  with $\sim 63\%$ percentage error for the FSRQS. The estimates seem to be dominated by random errors. For the variability Doppler factors, estimates are dominated by systematic errors, such as pileups, due to cadence observations which sets a limit on the fastest flare the survey is able to detect.  We estimated that the random error in the variability Doppler factor is $\sim 30\%$ making the variability method the most accurate method of the two, for sources where systematics are not a concern (for example, sources where the fastest flare is well-resolved and significantly longer in timescale than the cadence) in agreement with \cite{Lahteenmaki1999-II}.}

\item{The main limitation of the variability Doppler factor method assuming equipartition appears to be monitoring cadence. For this reason, long-term high-cadence blazar monitoring (such as the OVRO 15GHz monitoring program, \citealp{Richards2011}) can be an invaluable tool in deriving accurate Doppler factor estimates (within $30\%$ error) on a blazar-by-blazar basis.}

\item{Population models such as the one described here are unable to yield reliable and useful estimates for Doppler factors of individual blazars, and for this reason they should not be used in this fashion. }

\end{itemize}

In this work, we have used the Kolmogorov-Smirnov test to check whether the distributions of estimated  Doppler factors for blazar populations are inconsistent with our simulated ones. The reader should be cautioned that while inconsistency, when established at high significance, is statistically robust, a failure of the test to establish inconsistency could mean one of two things: either (a) the two distributions are consistent, or (b) it is impossible to tell because of small sample size (as is the case, for example, for BL Lacs in the Britzen et al sample.) For this reason, the physical interpretation of our results should be done with caution. An increased number of sources with Doppler factor estimates can strengthen our conclusions and further elucidate blazar physics.

\section*{Acknowledgments}

We are grateful to Talvikki Hovatta for insightful comments that improved this work. We would also like to thank Tony Readhead and Manolis Angelakis for valuable discussions.

This research was supported by the ``Aristeia'' Action of the  ``Operational Program Education and Lifelong Learning'' and is co-funded by the European Social Fund (ESF) and Greek National Resources, and by the European Commission Seventh Framework Program (FP7) through grants PCIG10-GA-2011-304001 ``JetPop'' and PIRSES-GA-2012-31578 ``EuroCal''.

\bibliographystyle{mnras}
\bibliography{bibliography} 

\label{lastpage}

\end{document}